\documentclass[12pt,preprint]{aastex}

\begin{document}

\title{Upper limit for $\gamma$-ray emission above $140\,\mathrm{GeV}$
  from the dwarf spheroidal galaxy Draco}

%
\author{
 J.~Albert\altaffilmark{a}, 
 E.~Aliu\altaffilmark{b}, 
 H.~Anderhub\altaffilmark{c}, 
 P.~Antoranz\altaffilmark{d}, 
 M.~Backes\altaffilmark{h},
 C.~Baixeras\altaffilmark{e}, 
 J.~A.~Barrio\altaffilmark{d},
 H.~Bartko\altaffilmark{f}, 
 D.~Bastieri\altaffilmark{g}, 
 J.~K.~Becker\altaffilmark{h}
% ...
}
 \altaffiltext{a} {Universit\"at W\"urzburg, D-97074 W\"urzburg, Germany}
 \altaffiltext{b} {IFAE, Edifici Cn., E-08193 Bellaterra (Barcelona), Spain}
 \altaffiltext{c} {ETH Zurich, CH-8093 Switzerland}
 \altaffiltext{d} {Universidad Complutense, E-28040 Madrid, Spain}
 \altaffiltext{e} {Universitat Aut\`onoma de Barcelona, E-08193 Bellaterra, Spain}
 \altaffiltext{f} {Max-Planck-Institut f\"ur Physik, D-80805 M\"unchen, Germany}
 \altaffiltext{g} {Universit\`a di Padova and INFN, I-35131 Padova, Italy}  
 \altaffiltext{h} {Universit\"at Dortmund, D-44227 Dortmund, Germany}
% ...

\begin{abstract}
The nearby dwarf spheroidal galaxy Draco with its high mass to light ratio 
is one of the most auspicious targets for indirect dark matter (DM)
searches. Annihilation of hypothetical DM particles can result in
high-energy $\gamma$-rays,  e.g. from
neutralino annihilation in the supersymmetric framework.\\
With the MAGIC telescope a search for a possible DM signal originating from
Draco was performed during 2007. The analysis of the data  results in a
flux upper limit ($2\sigma$) of  $1.1~\times~10^{-11}~\mathrm{photons}~\mathrm{cm}^{-2}~\mathrm{sec}^{-1}$
for photon energies above $140\,$GeV, assuming a point like
source. Furthermore, a comparison with predictions from supersymmetric models
is given. While our results do not constrain the mSUGRA phase parameter space,
a very high flux enhancement can be ruled out.
\end{abstract}


\section{Introduction}
Astronomical observations provide strong evidence for the existence 
of a new type of non-luminous, non-baryonic matter,
contributing to the total energy density of the universe about six times more
than baryonic matter \citep{Spergel2007}.
 This so-called Dark Matter (DM) makes its presence known through
 gravitational
effects, and could be made of so-far undetected relic 
particles from the Big Bang. Weakly Interacting Massive Particles (WIMP) 
are candidates for DM, with the lightest supersymmetric particle (neutralino)
 being one of the most favored candidates in the list of possible WIMPs.
 Stable neutralinos are predicted in many supersymmetric (SUSY) extensions of the
 Standard Model of Particle Physics~\citep{Jungman1995}.
Since the neutralinos are Majorana particles, pairs of neutralinos can
annihilate and produce Standard Model particles. Direct annihilations
into $\gamma\gamma$ or Z$\gamma$ produce a sharp line spectrum with a photon
energy depending on the neutralino mass. Unfortunately, these processes are
loop-suppressed and therefore very rare.
Neutralinos can also annihilate to  pairs of $\tau$ or quarks, leading
in subsequent  processes to $\pi^0$-decays, resulting in a continuous photon spectrum.\\
Draco is a dwarf spheroidal galaxy accompanying the Milky Way at a galactocentric 
distance of about $82\,$kpc. It is characterized by a high mass to light ratio  
$M/L > 200$, implying a high DM concentration~\citep{Mayer2007,Hooper2006} so complying
with the trend generally deduced for low-luminosity galaxies (e.g. \citet{Persic1996}).
\section{Expected $\gamma$-Ray Flux From Neutralino Self-Annihilation}

\begin{figure}[!hb] \centering
     \plotone{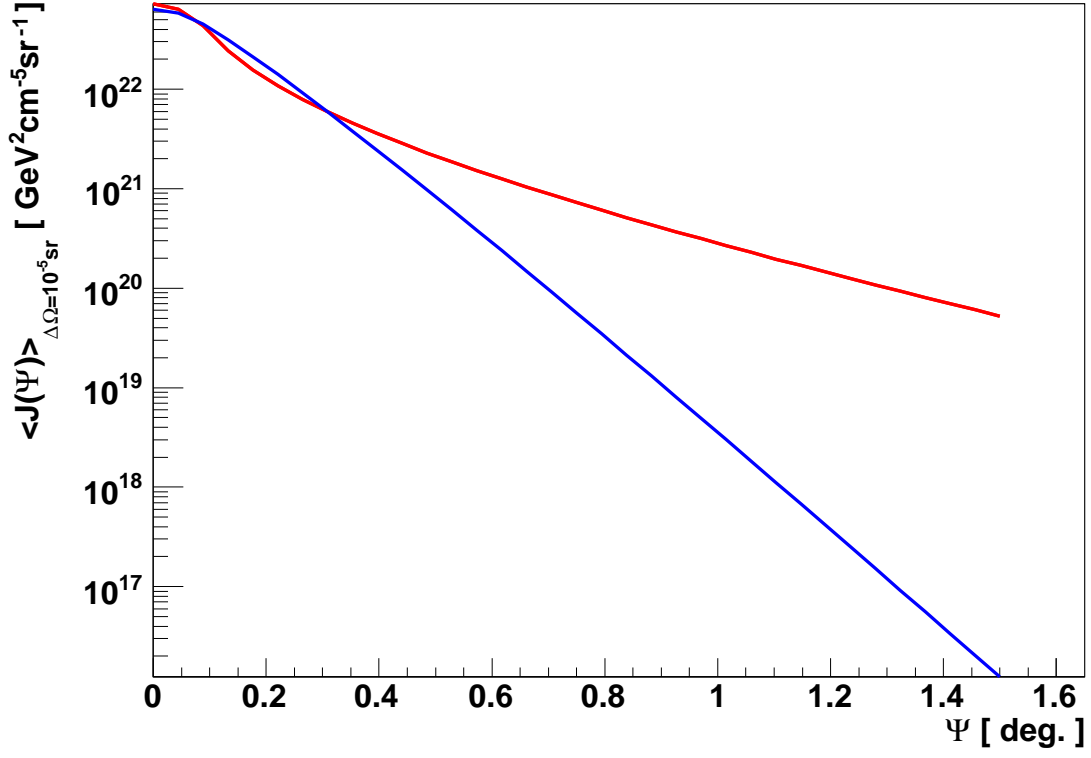}
   \caption{The factor $<J(\Psi)>_{\Delta\Omega}$ for the cusp (red)
     and the core (blue) profile. $\Psi=0$
     corresponds to the center of Draco. At an angular distance of $0.4^\circ$
     of the center of Draco,$<J(\Psi)>_{\Delta\Omega}$ is reduced by a factor of around 20
     for both models. }
  \label{Draco_figJPsi}
\end{figure}

The expected $\gamma$-ray flux depends on  details of the supersymmetric (SUSY) model as well as
on the density distribution of the DM in the observed
source. In general, the DM is assumed to be distributed in an extended halo
around spheroidal galaxies. The radial profile of the DM distribution in the
halo is modeled by a power law, $\rho_{\mathrm{DM}}(r)=Cr^{-\epsilon}$, where the
 parameter $\epsilon\ge 0$ describes the shape of the DM
 distribution in the crucial innermost region. $\epsilon=0$  results in a so-called {\it
   core model} with a central flat region, whereas profiles with
 $0.7<\epsilon < 1.2$ denote the so-called {\it cusp profiles}.
In addition, we chose an exponential cut-off as proposed by~\citet{Kazantzidis2004}:
\[\rho_{\mathrm{DM}}=Cr^{-\epsilon}\exp\left(-\frac{r}{r_b}\right)\]
with the values for $r_b$, $C$ and $\epsilon$ given in table 1 for a
cusp and a core profile for Draco. With the present angular resolution of the
MAGIC telescope ($0.1^\circ$), the two models
are indistinguishable due to the limited angular resolution of the
telescope which smears the determination of the profile (see figure \ref{Draco_figJPsi}).
From this figure we can see that the two profiles are discriminated at an
angular distance of $0.4^\circ$, where the intrinsic flux is already decreased by
a factor 20.
Depending on SUSY model parameters,
the annihilation cross section,  the average number of photons produced per annihilation and the
shape of the $\gamma$ spectrum can drastically change. Also a change in
the shape of
the DM density distribution along the line of sight can significantly change
the  $\gamma$-ray flux. Formula~(1) describes the expected $\gamma$-ray flux
above an energy $E_0$ from neutralino self-annihilation within Draco.
$\begin{array}[h]{lclr}
\Phi_{\gamma} (E>E_{0}) &=& \frac{1}{4\pi} f_{\mathrm{SUSY}}<J(\Psi)>_{\Delta\Omega}&(1)\\
\mathrm{where:}\\
f_\mathrm{SUSY}&=&\frac{N_{\gamma}(E>E_0)\left<\sigma v\right>}{2  m_{\chi}^{2}}\\
\rho(r) & &\textrm{DM density profile derived for Draco}\\
N_{\gamma} & &\textrm{photon yield per annihilation with } E >E_{0} \\
m_{\chi}& &\textrm{neutralino mass}\\
\left<\sigma v\right>& &\textrm{thermally averaged annihilation cross-section} \\
B(\Omega)& &\textrm{Point Spread Function (PSF) of the telescope}\\ 
\Psi& &\textrm{Pointing angle. } (\Psi=0\textrm{ for the center of Draco})\\
\Omega& &\textrm{solid angle of the telescope's resolution}\\
\textrm{los} & & \textrm{the line of sight}
\end{array}
$

\begin{figure}[!hb] \centering
     \plotone{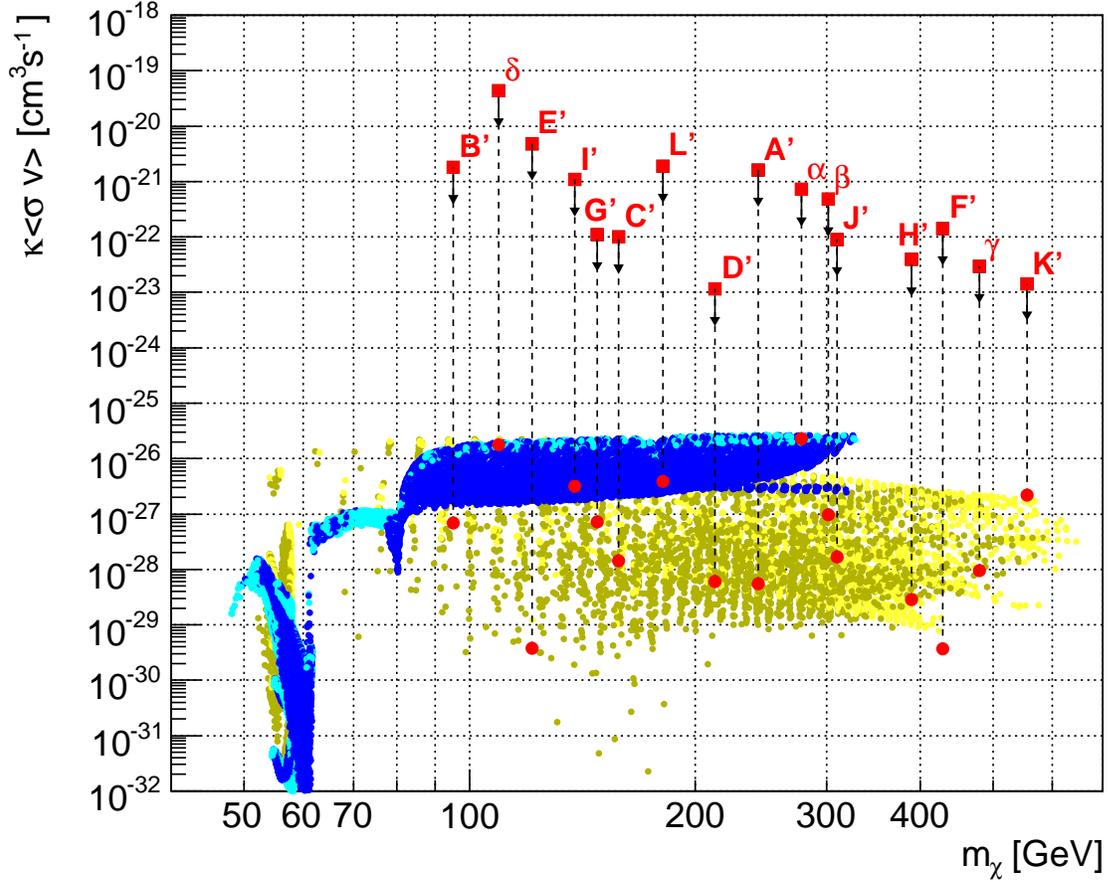}
   \caption{Thermally averaged neutralino annihilation cross section as a
     function of the neutralino mass
   for  mSUGRA models after renormalization to the relic density, as
   described in the text. The red dots marked with roman letters indicate benchmark models by
   \citep{Battaglia2003}. The ones marked with greek letters are models chosen
   by the authors.  The red boxes indicate the flux upper limit, displayed in units of
 $<\sigma v>$, assuming a smooth DM halo as given in table 1. See text for details.}

   \label{Draco_fig1}
\end{figure}

The factor $<J(\Psi)>_\Omega$ is shown for the cusp and the core model in figure \ref{Draco_figJPsi}. Even
though this factor converges for both models for small pointing angles $\Psi$
to the same value $<J(0)>_\Omega$,
there is an uncertainty in the distribution of the DM by
the existence of a hypothetical central black hole or a clumpy distribution of
the DM~\citep{Strigari2007,Colafrancesco2006}, which could lead to a
significant flux enhancement.\\
Due to the high predictive power of the mSUGRA framework, where the SUSY breaking effects are transmitted from
the high energy scale to the electroweak scale by the graviton,~\citep{Chamseddine1982, Inoue1982b, Inoue1982a, Inoue1983}
we simulated several million models
using the following parameters:  $m_0\le6$~TeV,
$m_{1/2}\le4$~TeV,
 -4~TeV $\le A_0 \le$ 4~TeV, tan$\beta\le 50$ and $\mu > 0$
\citep{Stark2005, Gondolo2000}. Figure \ref{Draco_fig1} summarizes the resulting
thermally averaged neutralino annihilation cross sections  for all
models not violating any observational 
constraints as well as
resulting in a total DM relic density $\Omega_{\mathrm{DM}}h^2$ in agreement with the $2\sigma$ upper limit
(u.l.) of 0.113 as derived from combined data from SPSS and
WMAP~\citep{Tegmark2006}.
The yellow points correspond to models with $m_0 \le$ 2 TeV (as favored by particle
physics), and the blue points represent $m_0 >$ 2 TeV.
Models resulting in a relic density below the lower WMAP-limit of 0.097 
are included, since neutralinos could  contribute only
a fraction to the total DM in the universe. For these models (shown as
dark blue points and dark yellow points in figure \ref{Draco_fig1}), a scale
factor of $\kappa =\left (\frac{\Omega_\chi h^2}{\Omega_{\mathrm{WMAP}}h^2}\right
)^2$ is applied to adjust for the DM relic density.\\


\section{Observation of Draco and Analysis}
Among all Imaging Air Cherenkov Telescopes in operation, MAGIC is the largest single-dish
facility (see e.g. \citep{MAGIC-commissioning, Cortina2005} for a
detailed description) and has the lowest energy threshold. MAGIC is located on the Canary
Island La Palma ($28.8^\circ$N, $17.8^\circ$W, 2200~m a.s.l.). The field of
view (FOV) of the 576-pixel photomultiplier camera is $3.5^\circ$. The angular resolution is $\sim0.1^\circ$ and the energy resolution
above 150~GeV is about 25\%. MAGIC has a trigger threshold of  $\sim$60~GeV for small zenith
angles (ZA), which increases for larger ZAs. \\
Data were taken in the false-source tracking (wobble) mode \citep{Fomin1994} with
two pointing directions at 24' distance and opposite sides of the source
direction in
May 2007 for a total observation time of 7.8~hours. Even though the source is
expected to be extended, the wobble mode is
justified, as at a distance of 24' of the center of Draco the expected flux
from this direction is less than 5\% of the flux coming from the center of Draco for
both the cusp and the core model (see figure \ref{Draco_figJPsi}). The ZA ranges
 between   $29^\circ$ and $42^\circ$. \\



\renewcommand{\arraystretch}{1.  }
Firstly, the calibration
of the data~\citep{Gaug2005} was performed. The arrival times
of the photons in core pixels ($>6$~photoelectrons (phe))  are required to be within a time window of 4.5~ns
and for boundary pixels ($>3$~phe) within a time window of 1.5~ns of a neighboring core pixel. The
next step includes the Hillas parameterization of the shower images
\citep{Hillas1985}. Two additional parameters, namely the time gradient along the
main shower axis and the time spread of the shower pixels, were computed \citep{Tescaro2007}.
Hadronic background suppression was achieved using the Random Forest (RF)
method~\citep{Breiman2001, Bock2007}, where for each
event the so-called \textsc{Hadronness} is computed, based on the Hillas
 and the time parameters. $\gamma$/hadron separation is realized by a cut in \textsc{Hadronness},  derived
from a $\gamma$ Monte Carlo (MC)  test sample \citep{Heck1998,Majumdar2005}, requiring a
$\gamma$-cut efficiency of 70\%. Moreover, the RF method was also
used for the energy estimation. \\

\section{Results}
\begin{figure}[t] \centering
     \plotone{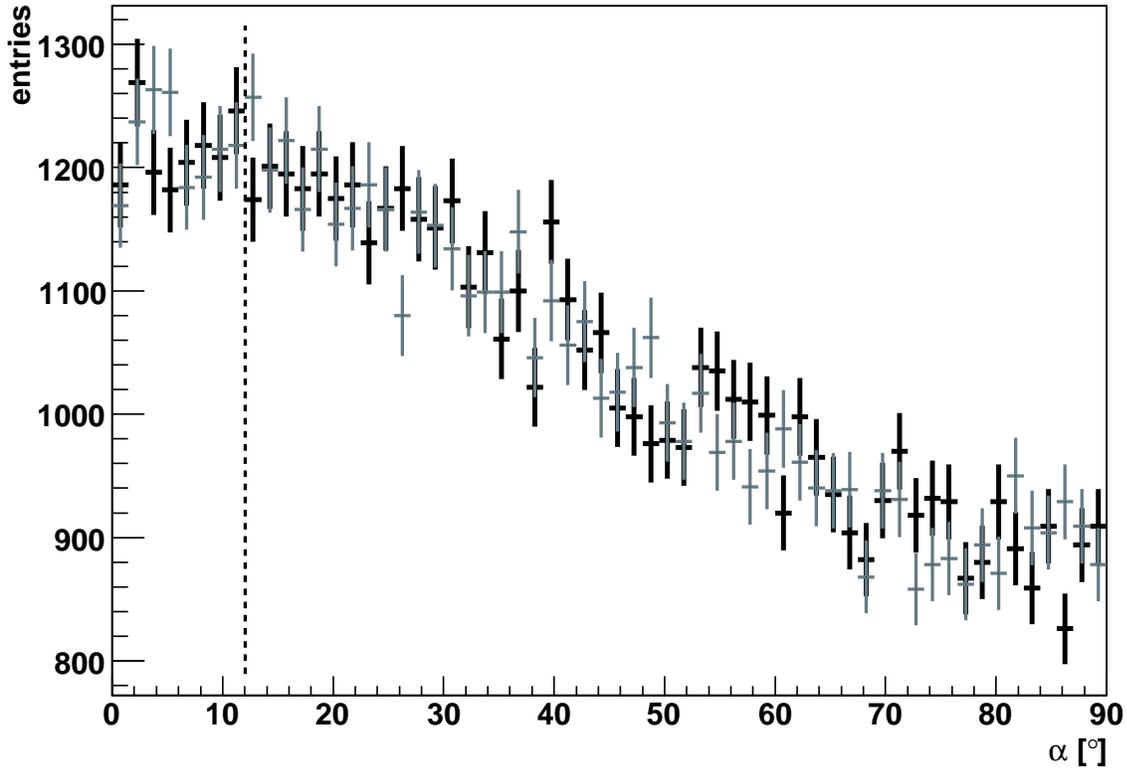}
   \caption{Distribution of the $\alpha$ parameter for $\gamma$-ray candidates coming
     from the center of Draco (black marker) and background
    (grey markers) for data taken between 05/09/2007 - 05/20/2007. 
The energy threshold is 140~GeV. For the signal region ($\alpha<12^\circ$),
are the number of ON events: 10883 and the number of OFF
events: 10996. With the method of \citep{Rolke2005}, the $2\sigma$ upper limit
on excess events is 231.}
   \label{alpha}
\end{figure}

We searched for a steady $\gamma$-ray emission from the direction of
the dwarf spheroidal galaxy Draco. The analysis energy threshold defined as
the peak of the energy distribution of MC generated $\gamma$ events after
cuts is 140~GeV. Images of
air showers initiated by $\gamma$-rays coming from the center of Draco are characterized by a
small 
$\alpha$ image parameter, which is the angle between the main axis of the
shower image and the connecting line between the center of gravity of the
shower and the source position in the camera.
The distribution of
$\alpha$ is shown in figure \ref{alpha} for all events after cuts. No significant excess was
found. The $2\sigma$
u.l. on the number of excess events was calculated using the method of
\citet{Rolke2005}, applying a systematic error of 30\%. The number of excess
events were converted into an integral flux u.l., 
depending on the assumed underlying spectrum. For a power law with spectral
index $-1.5$, typical for a DM annihilation spectrum, and assuming a point
like source, the $2\sigma$ u.l. is:
\[\Phi_{2\sigma}(E>140~\mathrm{GeV}) = 1.1\times 10^{-11}\,\mathrm{photons}\,\mathrm{cm}^{-2}\,\mathrm{s}^{-1}.\]

For different mSUGRA model parameters using the benchmark points defined in
\citep{Battaglia2003} and for other models, we computed the
$\gamma$-ray spectra expected from neutralino annihilations. 
Assuming these underlying spectra, the u.l. on the integrated flux above 140~GeV
is computed. Using formula (1) and assuming a  DM distribution following
the profiles according to table 1, the flux u.l. is displayed in the units of a thermally averaged
cross section in table 2 and in figure \ref{Draco_fig1}: 
\begin{eqnarray*}
\begin{array}{rl}

F_{2\sigma}=& \frac{\Phi_{2\sigma}(E>E_{0})}{\Phi_{\gamma} (E>E_{0})}<\sigma v>
\end{array}
\label{eq:flux}
\end{eqnarray*} 
As can be seen, the measured flux u.l. is several orders of magnitude larger
than predicted for the smooth DM distribution.
But a high clumpy structure of the DM distribution or a central black
hole could provide a significant flux enhancement \citep{Strigari2007,Colafrancesco2006}, which would
decrease $F_{2\sigma}$.  The analysis
presented here can set a limit on the flux enhancement depending on the mSUGRA
input parameters. For the benchmark models the
values for $\kappa <\sigma  v>$ and $F_{2\sigma}$ are displayed in table 2 and in
figure \ref{Draco_fig1}. For these models, the u.l. on the
flux enhancement is around $O(10^3~-~10^9)$.

\section{Conclusions}
We present the first search for $\gamma$-rays from the direction of Draco
using an Imaging Air Cherenkov Telescope (IACT). No signal was detected.
The $2\sigma$ u.l. on a steady $\gamma$-ray emission above $140\,\mathrm{GeV}$ originating from Draco does not
exceed $1.1\times 10^{-11}\,\mathrm{photons}\,\mathrm{cm}^{-2}\,\mathrm{s}^{-1}$ if
the underlying spectrum follows a power law with spectral index -1.5.\\
For the mSUGRA benchmark models defined in \citep{Battaglia2003} and assuming
a smooth DM density distribution for Draco as given in
\citep{Sanchez-Conde2007}, our flux upper limits are O($10^3$ - $10^9$) above
the predicted values. It is therefore not possible to constrain the mSUGRA
phase space by these results, but a very high flux enhancement can be
excluded. \\
Even though an indirect DM detection by measuring  $\gamma$-rays from neutralino
annihilation within the halo of Draco seems for present IACTs out of reach,
future satellite telescopes like GLAST with lower energy thresholds might be sensitive
enough to reach the mSUGRA parameter space.

\section{Acknowledgments}

We would
like to thank the IAC for the excellent working conditions at the
Observatory de los Muchachos in La Palma. The support of the
German BMBF and MPG, the Italian INFN and the Spanish CICYT is
gratefully acknowledged. This work was also supported by ETH
Research Grant TH~34/04~3 and the Polish MNiI Grant 1P03D01028.

\clearpage


\begin{thebibliography}{28}
\expandafter\ifx\csname natexlab\endcsname\relax\def\natexlab#1{#1}\fi

\bibitem[{{Albert} {et~al.}(2007)}]{Bock2007}
{Albert}, J. {et~al.} 2007, astro-ph/0709.3719

\bibitem[{Baixeras {et~al.}(2004)}]{MAGIC-commissioning}
Baixeras, C. {et~al.} 2004, Nucl. Instrum. Meth., A518, 188

\bibitem[{Battaglia {et~al.}(2004)}]{Battaglia2003}
Battaglia, M. {et~al.} 2004, Eur. Phys. J., C33, 273

\bibitem[{{Bergstr{\"o}m} \& {Hooper}(2006)}]{Hooper2006}
{Bergstr{\"o}m}, L. \& {Hooper}, D. 2006, \prd, 73, 063510

\bibitem[{Breiman(2001)}]{Breiman2001}
Breiman, L. 2001, Machine Learning, 45, 5

\bibitem[{Chamseddine {et~al.}(1982)Chamseddine, Arnowitt, \&
  Nath}]{Chamseddine1982}
Chamseddine, A.~H., Arnowitt, R., \& Nath, P. 1982, Phys. Rev. Lett., 49, 970

\bibitem[{{Colafrancesco} {et~al.}(2007){Colafrancesco}, {Profumo}, \&
  {Ullio}}]{Colafrancesco2006}
{Colafrancesco}, S., {Profumo}, S., \& {Ullio}, P. 2007, \prd, 75, 023513

\bibitem[{Cortina {et~al.}(2005)}]{Cortina2005}
Cortina, J. {et~al.} 2005, 5, 359, proc. of the 29th ICRC, Pune, India

\bibitem[{Fomin {et~al.}(1994)Fomin, Stepanian, A., Lamb, Lewis, Punch, \&
  Weekes}]{Fomin1994}
Fomin, V.~P., Stepanian, A., A., Lamb, R.~C., Lewis, D.~A., Punch, M., \&
  Weekes, T.~C. 1994, Astroparticle Physics, 2, 137

\bibitem[{Gaug(2005)}]{Gaug2005}
Gaug, M. 2005, proc. of the 7th Workshop on Towards a Network of Atmospheric
  Cherenkov Detectors 2005, Palaiseau, France, 27-29 Apr 2005

\bibitem[{Gondolo {et~al.}(2000)Gondolo, Edsjo, Bergstrom, Ullio, \&
  Baltz}]{Gondolo2000}
Gondolo, P., Edsjo, J., Bergstrom, L., Ullio, P., \& Baltz, E.~A. 2000,
  astro-ph/0012234

\bibitem[{Heck {et~al.}(1998)Heck, Schatz, Thouw, Knapp, \&
  Capdevielle}]{Heck1998}
Heck, D., Schatz, G., Thouw, T., Knapp, J., \& Capdevielle, J.~N. 1998,
  fZKA-6019

\bibitem[{Hillas(1985)}]{Hillas1985}
Hillas, A.~M. 1985, proc. of the 19th ICRC, La Jolla, 3, 445

\bibitem[{Inoue {et~al.}(1982{\natexlab{a}})Inoue, Kakuto, Komatsu, \&
  Takeshita}]{Inoue1982b}
Inoue, K., Kakuto, A., Komatsu, H., \& Takeshita, S. 1982{\natexlab{a}}, Prog.
  Theor. Phys., 68, 927

\bibitem[{Inoue {et~al.}(1982{\natexlab{b}})Inoue, Kakuto, Komatsu, \&
  Takeshita}]{Inoue1982a}
---. 1982{\natexlab{b}}, Prog. Theor. Phys., 67, 1889

\bibitem[{Inoue {et~al.}(1984)Inoue, Kakuto, Komatsu, \& Takeshita}]{Inoue1983}
---. 1984, Prog. Theor. Phys., 71, 413

\bibitem[{Jungman \& Kamionkowski(1995)}]{Jungman1995}
Jungman, G. \& Kamionkowski, M. 1995, Phys. Rev. D, 51, 3121

\bibitem[{{Kazantzidis} {et~al.}(2004){Kazantzidis}, {Mayer}, {Mastropietro},
  {Diemand}, {Stadel}, \& {Moore}}]{Kazantzidis2004}
{Kazantzidis}, S., {Mayer}, L., {Mastropietro}, C., {Diemand}, J., {Stadel},
  J., \& {Moore}, B. 2004, \apj, 608, 663

\bibitem[{Majumdar {et~al.}(2005)Majumdar, Moralejo, Bigongiari, Blanch, \&
  Sobczynska}]{Majumdar2005}
Majumdar, P., Moralejo, A., Bigongiari, C., Blanch, O., \& Sobczynska, D. 2005,
  5, 203, proc. of the 29th International Cosmic Ray Conference (ICRC 2005),
  Pune, India, 3-11 Aug 2005

\bibitem[{Mayer {et~al.}(2007)Mayer, Kazantzidis, Mastropietro, \&
  Wadsley}]{Mayer2007}
Mayer, L., Kazantzidis, S., Mastropietro, C., \& Wadsley, J. 2007, Nature, 445,
  738

\bibitem[{{Persic} {et~al.}(1996){Persic}, {Salucci}, \& {Stel}}]{Persic1996}
{Persic}, M., {Salucci}, P., \& {Stel}, F. 1996, \mnras, 281, 27

\bibitem[{Rolke {et~al.}(2005)Rolke, L\'opez, \& Conrad}]{Rolke2005}
Rolke, W.~A., L\'opez, A.~M., \& Conrad, J. 2005, Nucl. Instrum. and Meth.,
  A551, 493

\bibitem[{Sanchez-Conde {et~al.}(2007)}]{Sanchez-Conde2007}
Sanchez-Conde, M.~A. {et~al.} 2007, Phys. Rev. D 76, 123509

\bibitem[{{Spergel} {et~al.}(2007){Spergel}, {Bean}, {Dor{\'e}}, {Nolta},
  {Bennett}, {Dunkley}, {Hinshaw}, {Jarosik}, {Komatsu}, {Page}, {Peiris},
  {Verde}, {Halpern}, {Hill}, {Kogut}, {Limon}, {Meyer}, {Odegard}, {Tucker},
  {Weiland}, {Wollack}, \& {Wright}}]{Spergel2007}
{Spergel}, D.~N., {Bean}, R., {Dor{\'e}}, O., {Nolta}, M.~R., {Bennett}, C.~L.,
  {Dunkley}, J., {Hinshaw}, G., {Jarosik}, N., {Komatsu}, E., {Page}, L.,
  {Peiris}, H.~V., {Verde}, L., {Halpern}, M., {Hill}, R.~S., {Kogut}, A.,
  {Limon}, M., {Meyer}, S.~S., {Odegard}, N., {Tucker}, G.~S., {Weiland},
  J.~L., {Wollack}, E., \& {Wright}, E.~L. 2007, \apjs, 170, 377

\bibitem[{Stark {et~al.}(2005)Stark, H{\"a}fliger, Biland, \&
  Pauss}]{Stark2005}
Stark, L.~S., H{\"a}fliger, P., Biland, A., \& Pauss, F. 2005, JHEP, 08, 059

\bibitem[{{Strigari} {et~al.}(2007){Strigari}, {Koushiappas}, {Bullock}, \&
  {Kaplinghat}}]{Strigari2007}
{Strigari}, L.~E., {Koushiappas}, S.~M., {Bullock}, J.~S., \& {Kaplinghat}, M.
  2007, \prd, 75, 083526

\bibitem[{{Tegmark} {et~al.}(2006)}]{Tegmark2006}
{Tegmark}, M. {et~al.} 2006, \prd, 74, 123507

\bibitem[{Tescaro {et~al.}(2007)}]{Tescaro2007}
Tescaro, D. {et~al.} 2007, proc. of the 30th ICRC, Merida, Mexico

\end{thebibliography}

\begin{center}
\begin{table}[h]{\small\center
\begin{tabular}{l c c c}
 \hline
       profile &$ \epsilon$& $C$ & $r_b (\mathrm{kpc})$ \\     
 \hline
       cusp &1& $3.1 \times 10^7\,M_{\odot}\,\mathrm{kpc}^{-2}$ &1.189  \\
       core  &0& $3.6 \times 10^8\,M_{\odot}\,\mathrm{kpc}^{-3}$ & 0.238\\
 \hline
 \end{tabular}}
\label{dm_profile}
 \caption{Parameters considered for cusp  and  core DM
  density profiles~\citep{Sanchez-Conde2007}.}
\end{table}
\end{center}

\clearpage

\begin{table*}
  \rotate{
  \fontsize{7}{10} \selectfont
\begin{tabular}{c| c c c c c c  c cccccc }
 \hline
 model:  & A' & B' & C' & D' & E' & F' & G'&H'
 \\
 \hline
$m_0$ [GeV]  & 107 & 57 & 80 & 101 & 1532 & 3440 & 113 & 244\\
$A_0$ [GeV] & 0 & 0 & 0 & 0 & 0 & 0 & 0 & 0\\
$m_{1/2}$ [GeV] & 600 & 250 & 400 & 525 & 300 & 1000 & 375  & 935\\
$\tan\beta$  & 5 & 10 & 10 & 10 & 10 & 10 & 20 & 20\\
$m_\chi$ [GeV] & 243 & 95 & 158 & 212 & 121 & 428 & 148 & 389\\
 \hline
\begin{tabular}{c}
$\kappa<\sigma v>$\\
$[\mathrm{cm}^3\,\mathrm{s}^{-1}]$
\end{tabular}
 &  $5.55\cdot 10^{-29}$ &  $6.83\cdot 10^{-28}$ &  $1.42\cdot 10^{-28}$ &  $6.05\cdot 10^{-29}$ &  $3.74\cdot 10^{-30}$ &  $3.65\cdot 10^{-30}$ &  $7.18\cdot 10^{-28}$ &  $2.87\cdot 10^{-29}$\\
\begin{tabular}{c}
$F_{2\sigma}$\\
$[\mathrm{cm}^3\,\mathrm{s}^{-1}]$
\end{tabular}

 & $1.63\cdot 10^{-21}$ & $1.79\cdot 10^{-21}$ & $1.00\cdot 10^{-22}$ & $1.15\cdot 10^{-23}$ & $4.82\cdot 10^{-21}$ & $1.42\cdot 10^{-22}$ & $1.11\cdot 10^{-22}$ & $3.92\cdot 10^{-23}$\\
 \hline
\begin{tabular}{l}
ul on flux \\
 enhancement
\end{tabular}
 & $2.9\cdot 10^{7}$& $2.6\cdot 10^{6}$& $7.0\cdot 10^{5}$& $1.9\cdot 10^{5}$& $1.3\cdot 10^{9}$& $3.9\cdot 10^{7}$& $1.5\cdot 10^{5}$& $1.4\cdot 10^{6}$\\
 \hline
\end{tabular}

\begin{tabular}{c| c c c c c c  c cccccc }
 \hline
 model:   & I' & J' & K' & L' &  $\alpha$ & $\beta$ & $\gamma$ & $\delta$\\
 \hline
$m_0$  [GeV]& 181 & 299 & 1001 & 303&5980 & 180 & 1140 & 4540\\
$A_0$ [GeV] & 0 & 0 & 0 & 0&-300 & -2800 & -1800 & 300\\
$m_{1/2}$ [GeV]& 350 & 750 & 1300 & 450&680 & 720 & 1120 & 300\\
$\tan\beta$  & 35 & 35 & 46 & 47& 50 & 5 & 50 & 35\\
$m_\chi$  [GeV]& 138 & 309 & 554 & 181&277 & 301 & 479 & 109\\
 \hline
\begin{tabular}{c}
$\kappa<\sigma v>$\\
$[\mathrm{cm}^3\,\mathrm{s}^{-1}]$
\end{tabular}
 &  $3.17\cdot 10^{-27}$ &  $1.67\cdot 10^{-28}$ &  $2.22\cdot 10^{-27}$ &  $3.85\cdot 10^{-27}$&$2.27\cdot 10^{-26}$ &  $9.75\cdot 10^{-28}$ &  $9.43\cdot 10^{-29}$ &  $1.79\cdot 10^{-26}$\\
\begin{tabular}{c}
$F_{2\sigma}$\\
$[\mathrm{cm}^3\,\mathrm{s}^{-1}]$
\end{tabular}

 & $1.08\cdot 10^{-21}$ & $9.02\cdot 10^{-23}$ & $1.44\cdot 10^{-23}$ & $1.91\cdot 10^{-21}$&$7.22\cdot 10^{-22}$ & $4.78\cdot 10^{-22}$ & $2.97\cdot 10^{-23}$ & $4.39\cdot 10^{-20}$\\
 \hline
\begin{tabular}{l}
ul on flux \\
 enhancement
\end{tabular}
& $3.4\cdot 10^{5}$& $5.4\cdot 10^{5}$& $6.5\cdot 10^{3}$& $4.9\cdot 10^{5}$&$3.2\cdot 10^{4}$& $4.9\cdot 10^{5}$& $3.1\cdot 10^{5}$& $2.4\cdot 10^{6}$\\
 \hline
\end{tabular}
\caption{Thermally averaged neutralino annihilation cross section $<\sigma
  v>$, the u.l. on the flux $F_{2\sigma}$, displayed in units of  $<\sigma v>$,
  and the $2\sigma$ u.l. on the flux enhancement. Models A' - L' correspond to
  the benchmark models given by  \citep{Battaglia2003}. Models $\alpha$ -
  $\delta$ are typical models chosen by the authors with $A_0\neq 0$.}}
\end{table*}

\clearpage

\end{document}